\documentclass[11pt]{article}

\def\be{\begin{eqnarray}}
\def\ee{\end{eqnarray}}
\def\bq{\begin{equation}}
\def\eq{\end{equation}}
\def\ben{\begin{enumerate}}\def\een{\end{enumerate}}

\def\roughly#1{\mathrel{\raise.3ex\hbox{$#1$\kern-.75em%
\lower1ex\hbox{$\sim$}}}}

\usepackage{epsfig}

\begin{document}
\begin{titlepage}

\hfill FTUV-04-0125

\hfill IFIC-04-05

 \vspace{1.5cm}
\begin{center}
\ \\
{\bf\LARGE The scalar glueball spectrum}
\\
\vspace{0.7cm} {\bf\large Vicente Vento} \vskip 0.7cm

{\it  Departamento de F\'{\i}sica Te\'orica and Instituto de
F\'{\i}sica Corpuscular}

{\it Universidad de Valencia - Consejo Superior de Investigaciones
Cient\'{\i}ficas}

{\it 46100 Burjassot (Val\`encia), Spain, }

{\small Email: Vicente.Vento@uv.es}

\end{center}
\vskip 1cm \centerline{\bf Abstract} \vskip 0.3cm

We discuss scenarios for scalar glueballs using arguments based on
sum rules, spectral decomposition, the $\frac{1}{N_c}$
approximation, the scales of the strong interaction and the
topology of the flux tubes. We analyze the phenomenological
support of those scenarios and their observational implications.
Our investigations hint a rich low lying glueball spectrum.
\vspace{2cm}

\noindent Pacs: 12.38-t, 12.38.Aw,14.70.-e, 14.70.-j


\end{titlepage}

\section{Introduction}

\indent\indent The glueball spectrum has attracted much attention
since the formulation of the theory of the strong interactions
Quantum Chromodynamics
(QCD)\cite{FritzschGellMannLeutwyler,MinkowskiFritzsch}. QCD sum
rules \cite{ShifmanVainstein} and models \cite{Isgur,Petersen}
have been used to determine their spectra and properties. Lattice
QCD computations, both in the pure glue theory and in the quenched
approximation of QCD, have been used to determine their spectra
\cite{Weingarten,Teper}. It has become clear by now that it is
difficult to single out which states of the hadronic spectrum are
glueballs because we lack the necessary knowledge to determine
their decay properties \cite{Close}. Moreover the strong expected
mixing between glueballs and quark states leads to a broadening of
the possible glueball states which does not simplify their
isolation \cite{Narison1}. The wishful sharp resonances which
would confer the glueball spectra the beauty and richness of the
baryonic and mesonic spectra are lacking. This confusing picture
has led to a loss of theoretical and experimental interest in
these hadronic states. However, it is important to stress, that if
they were to exist they would be a beautiful and unique
consequence of QCD .

Glueballs have not been an easy subject to study and much debate
has been associated with their properties\cite{Narison1}. Even the
quantum numbers of the lowest lying glueball have not been agreed
upon until recently. There is now a general consensus that the
lightest glueball is a $0^{++}$ \cite{West}. However, its
properties, i.e., mass and widths still differ among the various
calculations. Dominguez and Paver \cite{dominguez}, Bordes,
Pe\~narrocha and Gim\'enez \cite{Bordes}, and Kisslinger and
Johnson \cite{Kisslinger} obtain by means of low energy theorems
and/or sum rule calculations with (or without) instanton
contributions a low lying (mass $<$ 700 MeV), narrow
($\Gamma_{\pi\pi} <$ 100 MeV) scalar glueball. Narison and
collaborators\cite{Narison2} using a two (substracted and
unsubstracted) sum rules prefer a broader (200-800 MeV), heavier
(700-1000 MeV) gluonium whose properties imply a strong violation
of the Okubo-Zweig-Ishimura's (OZI) rule \footnote{However, a
lighter glueball would be narrow since the coupling to $\pi \pi$
is proportional to the square of the mass.}. In a recent state of
the art sum rule calculation, Forkel \cite{forkel}, obtains the
gluonium at 1250 $\pm$ 200 MeV with a large width ($\sim$300 MeV).
However he has some strength at lower masses which he is not able
to ascribe to a resonance in the fits \footnote{Private
communication.}. Lattice QCD \cite{Weingarten,Teper} produce heavy
glueballs. Present day interpretation of
experiments\cite{AmslerClose,Barberis} claim a heavy glueball
($\sim$1500 MeV). We found illuminating the discussion of
Kisslinger and Johnson \cite{Kisslinger} since using their
calculation they can explain the existence of two scalar
glueballs, a light one ($\sim$500 MeV) and heavy one ($\sim$1700)
MeV, by studying the influence of the higher condensates in their
sum rule approach.

To investigate the scalar glueball sector we develop our
description initially in a world where the OZI rule is exactly
obeyed, i.e., decays into quarks which require gluons are strictly
forbidden. OZI dynamics (OZID) generates a glueballs spectrum
which is formed of towers of states disconnected from mesons,
baryons and leptons. The lowest lying scalar glueball (hereafter
called $g$) is, in this world, a bound state of two strongly
interacting gluons with a torus type flux tube topology
\cite{Hooftconfinement}. OZID confers this topology a Super
Selection rule inhibiting any decays from this state into other
particles. It is therefore stable and (almost) invisible since it
only interacts with other glueballs and gravitationally. $g$,
arises as a pseudo-Goldstone boson of broken scale invariance and
therefore its mass is provided by the gluon condensate.

However, OZID is an idealized scenario, which breaks down, and
through this breaking the interactions of the glueballs with
quarks, and through them with all other standard model probes,
arise. The implementation of this breaking leads to scenarios,
which we analyze.

\section{QCD scalars}

To transform the OZID scenario into a Gedanken picture of reality
we need the support of theory. Our basic assumption is that the
trace anomaly gives rise in QCD to a dilaton which is a
pseudo-Goldstone boson of scale invariance in line with the
arguments of anomaly cancellation of 't Hooft \cite{Hooftanomaly},
which have been so succesfully applied to the axial anomaly
\cite{Hooftanomaly,Frishman,Witten}. The would be dilaton will
describe the $0^{++}$ glueball ground state. In the extreme OZID
picture, gluodynamics is the theory describing $g$ and becomes
effectively

\bq L = \frac{1}{2}\;(\partial g)^2 + V(g) \label{g} \eq
where the potential $V(g)$ has been constructed to satisfy the
anomaly constraint and some low energy theorems
\cite{Schechter,Migdal,EllisLanik}. A consequence of this analysis
is the following relation between the mass of the dilaton and the
condensate,

\bq m_g^2 f_g^2 = - 4 \; <0| \; \frac{\beta(\alpha_s)}{4
\alpha_s}\; G^2\;|0>,\label{mgfg} \eq
where $f_g = <0|w|0>$  is the dilaton's vacuum expectation value,
$m_g$ the dilaton  mass, and the right hand side arises from the
anomaly. This relation was also obtained by Novikov et al.
\cite{Novikov} isolating the leading power correction in their
calculation\footnote{The validity of this approximation in their
scheme sets the limit on the mass of $g$, i.e. full consistency in
their analysis would imply a very small mass for $g$ and the
effective theory eq.(\ref{g}) would be a good realization of QCD
in this sector.}.

The theoretical support for OZID we find in the $\frac{1}{N_c}$
expansion of QCD. Eq.(\ref{mgfg}) is consistent with the expected
behavior

\bq m_g \sim 1 \;\;\; \mbox{and}\;\;\;  f_g \sim N_c.
\label{fg}\eq
Let us introduce the following correlator

\bq \Pi(q^2) = i\int dx e^{iqx} <0|\; T
\left(\frac{\beta(\alpha_s)}{4 \alpha_s} G^2(x)
\frac{\beta(\alpha_s)}{4 \alpha_s} G^2(0)\right)|0>. \label{pi}
\eq
It is known that \cite{Novikov}

\bq
\Pi (0) = - 4 <0|\frac{\beta(\alpha_s)}{4 \alpha_s} G^2(0)|0>,
\label{pi0}
\eq
which is related to the energy of the vacuum. To leading order in
$\frac{1}{N_c}$,

\bq
 \Pi (q^2) = \sum_{glueballs} \frac{N^2_c a_n^2}{M_n^2 -
q^2} + \sum_{mesons} \frac{N_c c_n^2}{m_n^2 - q^2}, \label{pi1/n}
\eq
where $M_n$ and $m_n$ represent respectively the masses of the
glueballs and mesons contributing to the correlator, and the
numerators are related to the following transition matrix elements

\bq
 N_c a_n = <0|\frac{\beta(\alpha_s)}{4 \alpha_s} G^2 |\;nth \;
glueball> \label{an} \eq
and

\bq
 \sqrt{N_c} c_n = <0|\frac{\beta(\alpha_s)}{4 \alpha_s} G^2 |
\;nth \; meson>. \label{cn} \eq
In the extreme $\frac{1}{N_c}$ limit at low $q^2$, $\Pi (q^2)$ is
dominatd by the lowest mass glueball $m_g$, and then

\bq |<0| \frac{\beta(\alpha_s)}{4 \alpha_s} G^2 | g >| ^2 = - 4
m_g^2 < 0 | \frac{\beta(\alpha_s)}{4 \alpha_s} G^2 | 0 >
\label{lowq} \eq
in agreement with the effective theory, i.e.

\bq m_g^2 f_g = < 0 |\frac{\beta(\alpha_s)}{4 \alpha_s} G^2 | g
>.
\label{transitiong}
\eq
The above statements contradict the results of Voloshin and
Zakharov \cite{Voloshin}, which require that the matrix elements
of their scalar gluonic operator with light mesons are not
negligible. To avoid this contradiction we have to include the
lowest lying scalar meson, which is one order down in
$\frac{1}{N_c}$.

If we extend the analysis to include the lowest lying meson, which
we call $\sigma$, we get

\bq m_\sigma^2 f_\sigma = < 0 | \frac{\beta(\alpha_s)}{4 \alpha_s}
G^2 | \sigma >, \label{transitionsigma} \eq
and from the general properties of the $\frac{1}{N_c}$ expansion
we obtain,

\bq  m_\sigma \sim 1 \;\;\; f_\sigma \sim \sqrt{N_c}
.\label{fs}\eq

Let us estimate the masses of these two states following the
analysis of Shifman \cite{Shifman} although adapting his
philosophy to the above scheme. To calculate the $0^{++}$ gluonia
mass he assumes that the gluonic resonances couple strongly to the
quark degrees of freedom in line with the arguments of Voloshin
and Zakharov \cite{Voloshin}. We proceed in the OZID (large $N_C$)
limit, i.e. $g$ does not couple to them but $\sigma$ does, and it
is therefore the latter which plays the role of saturating the
matrix elements.

The glueball contributes to the spectral function as

\bq Im \Pi (q^2) = \pi m_g^2 f_g^2 \delta (q^2 - m_g^2)
\label{spectralg} \eq
The $\sigma$ follows the discussion in ref. \cite{Shifman} since
it may decay into other mesons, i.e. 2$\pi$, 2$\eta$, 2$K$,
$\ldots$, thus we obtain

\bq m_g^2 f_g^2 + \frac{s_r^2}{8 \pi^2} = - 4 < 0 |
\frac{\beta(\alpha_s)}{4 \alpha_s} G^2 | 0>, \label{mass} \eq
where we have used $ m_\sigma^2 f_\sigma^2 = \frac{s_r^2}{8 \pi^2}
$. The additional term appearing from the existence of the $g$
implies a reduction of the masses with respect to the cited
analysis.

Let me use $\frac{1}{N_c}$ in here,

$$ \frac{m_g^2 f_g^2 }{m_\sigma^2 f_\sigma^2} = N_c.$$

Repeating the numerical estimate of ref. \cite{Shifman} for
massless quarks and for $N_c = 3$ we get

$$ m_{res} = 600 \mbox{MeV}$$

and therefore

$$ m_g \sim m_\sigma \sim 600 \mbox{MeV} $$

and

$$ f_g \sim \sqrt{3} f_\sigma .$$

For massive quarks the Shifman's estimates lead to

$$ m_g \sim m_\sigma \sim 750 \mbox{MeV} $$

Thus $g$ and $\sigma$ have similar masses in this naive scenario.

In gluodynamics the absence of quarks increases the coupling
constant. The right hand side of Eq.(\ref{mass}) is related to the
energy of the vacuum which increases and the left hand side has no
contribution from the quarks, i.e. $\sigma$ meson, thus

\bq (m_g f_g)^{gluodynamics} \sim \sqrt{\frac{44}{27}} (m_g
f_g)^{QCD}. \label{mgfggluodynamics} \eq
Moreover we expect $f_g$ to decrease in gluodynamics since decay
channels into photons will be closed. This statement together with
our previous estimate Eq.(\ref{mgfggluodynamics}) leads to

$$ m_g^{gluodynamics} >  1  \mbox{GeV}. $$

This qualitative argument explains the mechanisms by which the
pure gauge calculations sees higher masses.

However, in line with the arguments of Kisslinger and Johnson
\cite{Kisslinger}, we should expect in gluodynamics two $0^{++}$
glueballs, separated by approximately 1 GeV and which would move
lower in energy once the effects of quarks are introduced.

\section{Topology and dynamics}

Nature does not realize OZID, namely the number of colors is not
very large. We have to establish a scheme for breaking OZID. How
should we incorporate corrections to the leading order in the
$\frac{1}{N_c}$ expansion? In order to understand how nature
departs from OZID we resort to symmetry breaking and topological
arguments.

Dynamical transmutation in $QCD$ gives rise to the confinement
scale, $\Lambda$, which introduces dimensions into a dimensionless
(apart from quark masses) theory. Conventional low energy physics
is governed by the chiral symmetry breaking scale $f_\pi$
\cite{Leutwyler}, which ultimately should be a function of
$\Lambda$. In our case low energy dynamics will be governed by
$f_g$ and $f_\sigma$ respectively. Recalling the results of
previous section we notice that $f_\pi \sim f_\sigma \sim
O(\sqrt{N_c})$, Eq. (\ref{fs}), while $f_g \sim O(N_c)$.
Eq.(\ref{fg}). The breaking of OZID is governed by powers of their
inverses. Thus we expect the corrections to the mesons to be
$O(\frac{1}{N_c})$ while that for the glueball
$O(\frac{1}{N_c^2})$. OZID is better realized in the glueball
sector than in the meson sector.

A second idea which guides our intuition about the breaking of
OZID is the topology of the flux tubes and their relation with
perturbative emission. The mesonic $q\bar{q}$ states have an
elongated, almost linear structure in their flux
tubes\cite{Bali,Trottier,Giacomo}. The glueballs in most
treatments arise from twisted flux tube configurations
\cite{Faddeev,Iwasaki,Buniy,Apreda}. In particular we conjecture
based on a the simplest possible non linear topology, namely a
torus like configuration \cite{Hooftconfinement}, the behavior of
particle emission.

Gluon and quark emission occur inside the flux tubes and therefore
the scale of the perturbative emission is limited by the
confinement size, i.e. the running coupling constant takes its
maximum possible value when the particles are emitted with the
lowest possible momentum, which is bounded from below by
Heisenberg's principle,

\begin{itemize}

\item[i)] for the meson: $ L < L_{conf} \sim
\frac{1}{\Lambda_{QCD}}
\sim 1 \mbox{fm}$.

\item[ii)] for the glueball\footnote{The torus flux tube is
basically a planar figure since the cross section radius of the
tube is small with respect to the other radius. Thus $\Delta p_x
\sim \frac{1}{2R}$ where $R$ is the radius of the large circle of
the torus. Thus $\Delta p = \sqrt{(\Delta p_x)^2 + (\Delta p_y)^2}
= \frac{1}{\sqrt{2}R}$. But $L_{conf} = 2\pi R$ thus $L <
\frac{L_{conf}}{\sqrt{2}\pi}$.}: $ L <
\frac{L_{conf}}{\sqrt{2}\pi} \sim \frac{1}{4} \mbox{fm}$.

\end{itemize}

Therefore,

\bq \alpha_{meson} \sim \alpha (L_{conf}) >> \alpha_{glueball} =
\alpha(\frac{L_{conf}}{\sqrt{2}\pi}) \label{alphas}\eq
where $\alpha$ is the running coupling constant.

This argument also suggests that OZID dynamics is a better
approximation in the case of glueballs than in the case of mesons
since for the former the perturbative emission is weak. We expect
therefore that the pure perturbative emission approximation of QCD
to gluon and quark emission for $g$ is very appropriate at any
scale, while for the $\sigma$ non perturbative chiral effects will
be important \cite{pich}.

\section{$g$-$\sigma$ Mixing}

Since $g$ and $\sigma$ have the same quantum numbers they can
easily mix in broken OZID and the observed particles are coherent
superpositions of them. We use the discussion of previous section
and $\frac{1}{N_c}$ arguments presented in the Appendix to
construct the breaking pattern.

We consider that $g$ and $\sigma$ mixed due to additional terms in
the hamiltonian which are of higher order in $\frac{1}{N_c}$.
Since $f_\sigma \sim \sqrt{N_c}$ and $f_g \sim N_c$ the following
is the most general hamiltonian in this reduced Fock space,
\bq \left(
\begin{array}{c c}
m & \delta \\
\delta & m +\Delta m \\
\end{array} \right)
\eq
where $\Delta m \sim \frac{1}{N_c}$, $\delta \sim
(\frac{1}{N_c})^{\frac{3}{2}}$ and we exclude terms
$O((\frac{1}{N_c})^2$ and higher powers. The diagonal basis of
this hamiltonian can be presented as,
\be \tilde{g} & =  & g \cos{(\theta/2)} - \sigma \sin{(\theta/2)}
,
\\ \tilde{\sigma} & =  & g \sin{(\theta/2)}  + \sigma \cos{(\theta/2)},
\ee
where the tilde labels the physical particles and $\theta$ is the
mixing angle.

\begin{figure}[htb]
\centerline{\epsfig{file=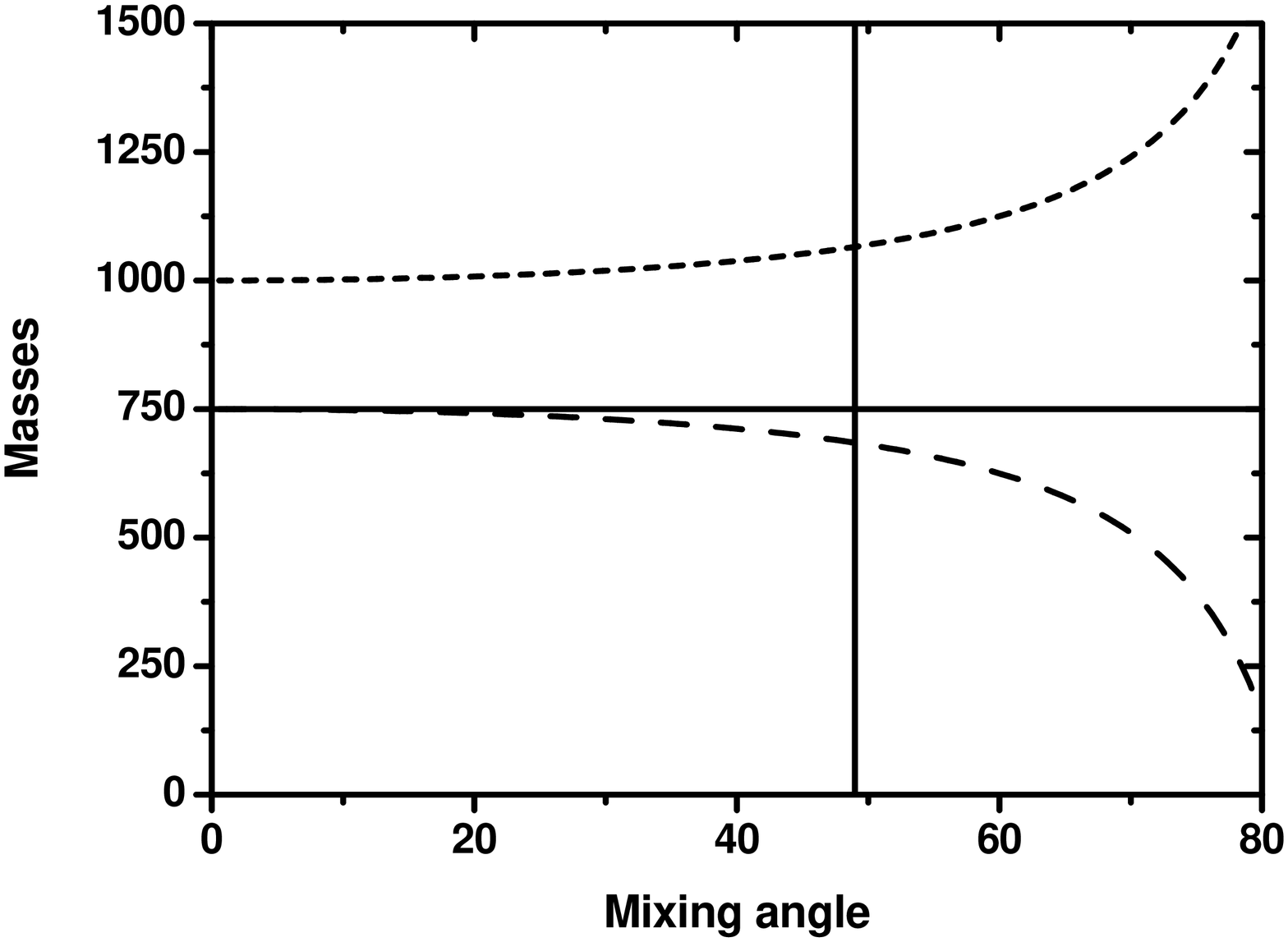,width=5.9cm,angle=270}
\hspace{0.5cm} \epsfig{file=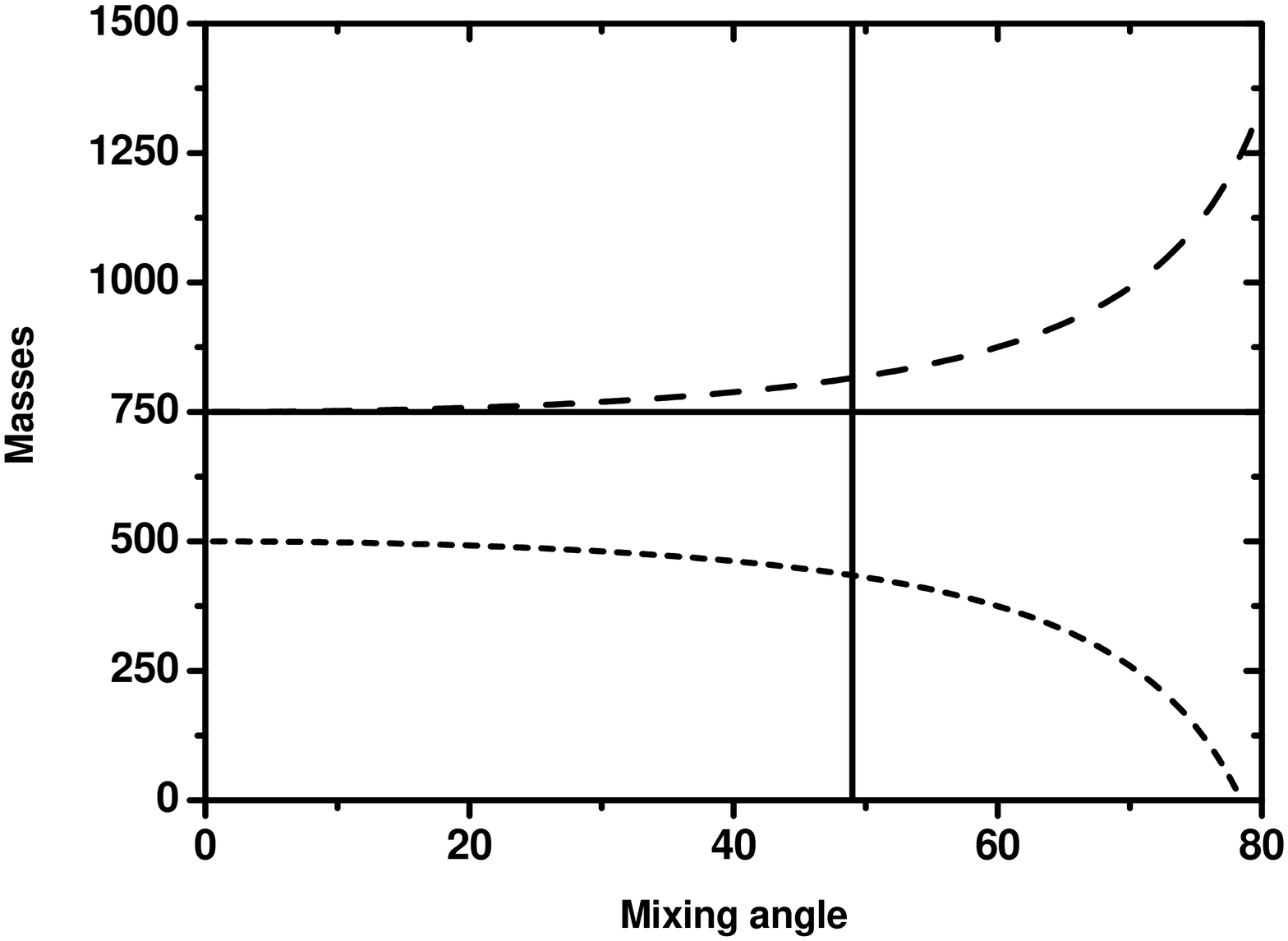,width=5.9cm,angle=270}}
\caption{The limiting values for the masses of the physical
${\tilde{g}}$ (solid--dashed lines) and  ${\tilde{\sigma}}$
(solid--short-dashed lines) are shown as a function of the mixing
angle for the range $0<|\Delta m| < 250$ MeV . The left (right)
figure corresponds to positive (negative) $\Delta m$. The
degenerate initial mass has been taken, as discussed in the text,
at $m = 750$ MeV. The vertical line defines the approximate limit
of the validity of the $\frac{1}{N_c}$ expansion.}
\label{2statemixmass}\vskip 1cm
\end{figure}

The masses of the physical particles become
\be m_{\tilde{g}}& = &m + \frac{\Delta m}{2} - r \nonumber\\
m_{\tilde{\sigma}}& = &m + \frac{\Delta m}{2} + r \\
\label{masses} \ee
where
\bq \tan{\theta} = \frac{2 \delta}{\Delta m} \eq
and
\bq r = \frac{\Delta m}{2} \sqrt{ 1 + (\frac{2\delta}{\Delta
m})^2} = \frac{\Delta m}{2 \cos{\theta}} \eq
In Figs.\ref{2statemixmass} we represent the masses of the
physical states as a function of the mixing angle $\theta$. The
curves separate possible mass regions. The horizontal solid line
represents the mass of both states for $\delta m \rightarrow
0$\footnote{We take a value for $\Delta m$ small enough so that
the deviation from this line which occurs for $\theta \rightarrow
\frac{\pi}{2}$, which leads ultimately to a $\pm \delta$ splitting
for $\delta$ finite, is beyond the shown values.}. The dashed
lines represent the value of $m_{\tilde{g}}$ for $\Delta m = \pm
250 MeV$ and the short-dotted lines those for
$m_{\tilde{\sigma}}$. To the left of the vertical line the values
are consistent with the $\frac{1}{N_c}$ expansion, the condition
that defines that line is
$$\tan{\theta} = \frac{2 \delta}{\Delta m} \sim \frac{2}{N_c} \sim
\frac{2}{3} $$.


In Figs.\ref{2statemixmass} the curves on the left show that the
two state mixing scenario for positive $\Delta m$ leads to a
"light" glueball with a mass in the range $650 $ MeV $ <
m_{\tilde{g}} < 750$ MeV and a scalar meson with a mass in the
range $750$ MeV$ < m_{\tilde{\sigma}} < 1050$ MeV . The
$\frac{1}{N_c}$ expansion favors small mixings in the physical
states. The curves on the right show that for negative $\Delta m$
the meson becomes lighter $450 $ MeV $ < m_{\tilde{\sigma}} < 750$
MeV, while the glueball becomes heavier $750$ MeV$ < m_{\tilde{g}}
< 850$ MeV.

 Let us speculate about strong OZID breaking. If we abandon
the $\frac{1}{N_c}$ expansion, i.e. allow the mixing matrix
elements to be larger than required by this approximation, the
masses separate notoriously and in particular the glueball (meson)
becomes very light in the $\Delta m > 0$ ( $\Delta m < 0$ )
scenario. Correspondingly, the associated meson (glueball) becomes
heavy. In this case however the mixing is large thus it is
difficult to talk about glueball or meson since both states are an
almost perfect mixture,i.e.  the $\tilde{g}$ state has a large a
large $\sigma$ component and the $\tilde{\sigma}$ state a large
glueball component.

We have performed a mathematical analysis of our theoretical
scheme, in the next section we put the present analysis under the
scrutiny of data.


\section{Discussion}

The OZID glueball does not interact with quarks, neither with
leptons nor electroweak gauge bosons, therefore in our approach it
is sterile. However, the physical glueball does because of its
admixture with the $\sigma$. From now on we will only talk about
the physical particles and we omit their tilde in the notation.
Using a $\sigma$-model interaction we get
\be  \Gamma_{\sigma\rightarrow 2\pi}& =& \frac{3}{64\pi f_\pi^2}
\left(\frac{m_\sigma^2 -m_\pi^2}{m_\sigma}\right)^2
\sqrt{m_\sigma^2 - 4m_\pi^2}\nonumber \\
& \sim & \frac{3}{64\pi}\frac{m_\sigma^3}{f_\pi^2}\sim 1.5
\left(\frac{m_\sigma (\mbox{GeV})}{1\mbox{GeV}}\right)^3
\mbox{GeV} \label{2piwidth},\ee
where we have taken $f_\pi \sim 100$ MeV and neglected terms
O($m_\pi^2$) in the last line.

Let us look at the lower spectrum of scalars shown in table 1.
Below $750$ MeV the only existing resonance is the broad
$f_0(600)$, whose mass and width are still quite undetermined.
Using the data on the width and using Eq.(\ref{2piwidth}) we
obtain

\bq 737 \mbox{MeV} < m_\sigma < 874 \mbox{MeV}. \eq

Thus the $\Delta m < 0$ scenario is discarded by the data.
Therefore the glueball is lighter than the meson, i.e. within the
limits of the $\frac{1}{N_c}$ expansion

\bq 650 \mbox{MeV} < m_\sigma < 750 \mbox{MeV}. \eq
Note that in this approximation the mixing angle is small and
therefore

\be  \Gamma_{g\rightarrow 2\pi} & \sim & 1.5 \sin ^2 {(\theta/2)}
\left(\frac{m_g (\mbox{GeV})}{1\mbox{GeV}}\right)^3 \mbox{GeV} <
100 \mbox{MeV},
\\
\Gamma_{\sigma\rightarrow 2\pi} & \sim &  1.5 \cos ^2 {(\theta/2)}
\left(\frac{m_\sigma (\mbox{GeV})}{1\mbox{GeV}}\right)^3
\mbox{GeV} > 500 \mbox{MeV}.\ee

Our analysis supports that the broad $f_0 (600)$ hides, within its
experimental indetermination our two states, the conventional
$\sigma$ meson and the lightest searched for glueball.

If we relax the OZID hypothesis we could arrive to an exotic
scenario in which for large mixings one of the states could have a
small mass close to the 2$\pi$ threshold and an extremely small
width due to the kinematical threshold factor appearing in
Eq.(\ref{2piwidth}). This exotic scenario would be characterized
by a quasi stable state close to the observed lower mass limit
($\sim 400$ MeV) and a broad width state in the upper mass limit
($\sim 1200$ MeV).

\begin{table}[tbp] \centering
\begin{tabular}
[c]{|c|c|c|c|c|c|}\hline  & $f_0(600)$& $ f_0(980) $ & $f_0(1370)
$ & $f_0(1500)$ & $f_0(1710)$\\ \hline mass (MeV) & $400 - 1200$ &
$980\pm 10$  & $ 120 - 1500$ &$ 1507 \pm 5$&$ 1714 \pm 5$\\
\hline  width (MeV) &$ 600 - 1000$
& $40 - 100$ & $200 - 500$ & $109 \pm 7$ & $140 \pm 10$\\
\hline  Decay &$ \pi \pi $ dominant &$\pi \pi$
dominant & $ \pi\pi$ seen & $\pi \pi$ 35\% & $ \pi \pi$ seen\\
modes &$ \gamma \gamma$ seen &$ K \bar{K}$ seen & $ 4\pi $ &
$4\pi$ 50\%& $ K \bar{K}$ seen \\  &$ $ &$\gamma \gamma$ seen &...
$\rho \rho $  dominant  & $\eta \eta$ 5\%& $\eta \eta$ seen\\  &$
$ &$$ & ... other $4\pi $ seen & $\eta \eta^{\prime}$ 2\%& $
 $\\ &$ $ &$$ & $\eta \eta $ seen  & $ K \bar{K}$ 9\% &
$$\\  &$ $ &$$ & $ K \bar{K}$ seen& $\gamma \gamma$ not seen&
$$\\ & $$ & $$ & $\gamma \gamma$ seen & $ $ & $$\\  \hline
\end{tabular}
\caption{The scalar spectrum according to the Particle Data Group
\cite{PDG}} \label{Table}
\end{table}

The $f_0(980)$, which belongs to the meson nonet, is too narrow to
correspond to our sigma-model state, since it survives the large
$N_c$ limit \cite{pich}, we ascribe it to the first mesonic
excitation. Since its width is relatively low it does not seem to
arise from a mixing with the lower lying states and therefore it
sets the upper bound for the mass of the lowest lying
$\sigma$-meson. Thus the existence of the $f_0(980)$ excludes, in
our view, the extreme exotic scenario and validates an approximate
OZID scheme.

The $f_0(1370)$ region is again ill-determined experimentally. In
this case new  channels, like $2 \eta$, open up. The same mass
analysis for the excitations  could be carried out, which would
lead us to conclude that two excited states, a glueball and a
meson, exist. Here we should not apply our naive sigma model width
and therefore our discussion for the widths is absent. However,
the recent analysis of Forkel\cite{forkel} concludes with the
existence of a broad glueball at $1250$ which corresponds to the
region around the upper mass limit. We conclude from this analysis
that in this region of the spectrum the $\Delta m < 0$ scenario
takes place and that the companion meson should have lower mass
than the glueball. The proximity of the $f_0(980)$, and a minimal
population hypothesis, leads us to propose that the $f_0(980)$ is
the required companion. The fact that the lower mass particle, a
meson in this case, is narrower also confirms the breaking scheme.

Finally if the $f_0(1500)$ is a glueball
\cite{AmslerClose,Barberis}, by assuming the same analysis, a new
$\Delta m > 0$ scenario may take places, which would ascribe the
$f_0(1700)$ as its companion meson. The $f_0(1500)$ could be also
the higher lying glueball of ref. \cite{Kisslinger} after mixing.

Thus, our analysis  leads to the existence of three glueball
states in the low lying scalar spectrum with three companion
mesons. The precise dynamical mechanisms by which they arise are
as of yet unknown, however more precise studies within the large
$N_c$ approximation might shed light to our proposal. The duality
of the $\Delta m$ mechanism which leads to an ordering of the
spectrum in the form
$$m_{g} < m_{\sigma}  < m_{\sigma_1} < m_{g_1} < m_{g_2} <
m_{\sigma_2} ....$$
has been guided by observation and physical intuition as explained
above.

The analysis could be completed by studying other decay modes. In
particular $2\gamma$ decays also hint about the mass orderings.
Using the trace anomaly \cite{EllisLanik}

\bq \Gamma_{\sigma\rightarrow 2\gamma} = \frac{\alpha^2}{16\pi^3}
\frac{m_{\sigma}^3}{{f_\sigma}^2} \sim 10.5 \left(
\frac{m_{\sigma}}{1\mbox{GeV}}\right)^3 \mbox{eV} \eq
where we have used $N_c = 3$ and $f_\sigma \sim f_ \pi \sim 100
\mbox{MeV}$. We obtain therefore
\be  \Gamma_{g \rightarrow 2\gamma} & \sim & 10.5 \sin ^2
{(\theta/2)} \left(\frac{m_g (\mbox{GeV})}{1\mbox{GeV}}\right)^3
\mbox{eV} < 1 \mbox{eV},
\\
\Gamma_{\sigma\rightarrow 2\gamma} & \sim &  10.5 \cos ^2
{(\theta/2)} \left(\frac{m_\sigma
(\mbox{GeV})}{1\mbox{GeV}}\right)^3 \mbox{eV} > 3 \mbox{eV}.\ee
Thus, in this weak mixing scenario, the glueball state is narrower
than the meson state.

\section{Conclusions}

We have analyzed the possible existence of a $0^{++}$ glueball low
lying state from different perspectives. The analysis has been
modelled by $\frac{1}{N_c}$ physics on which we have also based
our estimates. We are led to a scenario of weak OZID breaking and
a low mass glueball. This glueball is narrow since only its
$\sigma$ state component is allowed to decay and the small mixing
angle inhibits decays. It represents a beautiful example of OZID
dynamics.

The discussion and mechanisms can be repeated for the higher lying
scalars and a spectrum arises in which glueballs and scalar mesons
appear in pairs, with masses ordered according to the sign of the
$1/N_c$ breaking parameter $\Delta m$.

The lowest lying states, $g$ and $\sigma$, appear within the
$f_0(600)$ peak in agreement with previous estimates
\cite{Bordes,Narison2,Kisslinger}, and therefore they might be
difficult to isolate \cite{pennington}, although their widths are
vastly different for strong and electromagnetic decays. Maybe more
precise experiments could manage to see the two peaks.

The existence of low lying glueballs might strongly influence the
transition towards the Quark Gluon Plasma \cite{qgp,vento} and it
might be in this physical regime where it might appear
unquestioned.

The present investigation lies at the foundations for the
understanding of the scalar spectrum. Our reasonings can be made
more quantitative by lattice studies and more sophisticated model
studies. It opens up the possibility of understanding glueballs
and their dynamics.

\subsection*{Acknowledgments}

Correspondence with C.A. Dominguez,  M. Pennington and M. Teper is
gratefully acknowledged. The author has benefitted from long and
fruitful discussions with H. Forkel. This work was supported by
MCYT-FIS2004-05616-C02-01 and GV-GRUPOS03/094. The author
acknowledges the hospitality of CERN-TH where this work was
completed.

\end{document}